\documentclass[intlimits,twoside,a4paper]{article}

\usepackage[cp1251]{inputenc}

\usepackage{cmpj3}
\usepackage{graphicx}        

\newcommand{\be}{\begin{equation}}
	\newcommand{\ee}{\end{equation}}
\newcommand{\bea}{\begin{eqnarray}}
	\newcommand{\eea}{\end{eqnarray}}
\articletype{A part of the Special collection to the 100th anniversary of the birth of Ihor Yukhnovskii}

\issue{2026}{29}{3}{33501}
\doinumber{10.5488/CMP.29.33501}
\title[New applications of the work by Yukhnovskii and Kelbg II]{On the statistical theory of strong electrolytes and high-temperature
	plasmas: New applications of the work by Yukhnovskii and Kelbg II}

\author[W. Ebeling, M. Holovko]{W. Ebeling\orcid{0000-0003-0740-3016}\refaddr{label1}\thanks{Corresponding author:\email{ebeling@physik.hu-berlin.de}.},  M. Holovko\orcid{0000-0001-8114-5356}\refaddr{label2}}
\addresses{
	\addr{label1}{Institute of Physics, Humboldt University, Berlin, Germany}
	\addr{label2}{Yukhnovskii Institute for Condensed Matter Physics of the National Academy of Sciences of Ukraine, 1~Svientsitskii Str., 79011, Lviv, Ukraine}}

\Keywords{strong electrolytes, high-temperature plasmas, oscillatory correlations}
\date{Received 31 July 2026; revised 27 August 2026; accepted 31 August 2026; published 28 September 2026}

\begin{document}

\maketitle

\begin{abstract}
Exponential {potentials} were used since Kramers, Hellman, Glauberman, Yukhnovskii and Kelbg for
solving problems of quantum chemistry, ionic solutions and plasmas. Here we develop
the theory further and add new results, in particular with respect to the quantum-statistical theory.
In particular, we derive the Kelbg quantum potential for the exponential interactions  
and discuss screening as well as the problem of thermodynamic stability.
Further we give new applications to a qualitative theory of alkali plasmas and fusion 
plasmas. We show that the exponential potential and the early analytical results of Yukhnovskii and Kelbg allow 
a qualtative analytical treatment of such difficult problems as the phase transitions in electrolytes and alkali plasmas. 
\printkeywords
 \end{abstract}

\section{About exponential interactions}
\label{intro}
For modelling Coulomb-like systems, the class of exponential
potentials is well suited as demonstrated first by Kramers, Hellman, Glauberman, Yukhnovskii 
and Kelbg for solving problems of quantum chemistry, ionic solutions,  alkali plasmas and other interesting physical systems \cite{Hellmann1,Hellmann2,Hellmann3,GlaYuk52_1,Yukhn54,Kelbg59,Kelbg62,FalkenhagenEb71,Falkenhagen,Krasko,EbFoFi17,EbFoFi20,Yukhn80,Yukhn25}.
We develop further a systematic continuation of the work by Yukhnovskii and Kelbg on the exponential potential, which we started in \cite{EbHoCMP26}. We derive some new results  by using the 
known classical and quantum statistical theory, which might be useful for further applications. In particular, we develop new applications to the thermodynamics of alkali plasmas and dense high-temperature plasmas 
as in the center of stars or probably  in some future fusion devices as well.
The exponential potential as a regularized Coulomb potential proposed first in 1927 by Kramers in quantum theory, turned out to be very useful for applications in quantum chemistry \cite{Hellmann1,Hellmann2,Hellmann3}, in the theory of ionic solutions \cite{GlaYuk52_1,FalkenhagenEb71,Falkenhagen,Yukhn80} and for modelling alkali plasmas~\cite{Hansen,Richert}. The original form of the exponential potential contains only one free parameter $\alpha$ which is a reciprocal length, responsible for a regularization at $r = 0$ and is given by 
\bea
V_{ij} (r) = \frac{e_i e_j}{4 \piup  \epsilon r} [1 - \exp(-\alpha r)], \qquad \epsilon = \epsilon_0 \epsilon_r,
\eea
where $e_i$ are the ionic charges and $\epsilon_r$ is the relative dielectric constant of the medium. Glauberman and Yukhnovskii used this potential  first for 
solving equilibrium problems of electrolyte theory. The applications to electrolytes were extended by G\"unter Kelbg  to transport properties~{\cite{Kelbg59,Kelbg62}}. Afterwards Kelbg opened a completely new field by fruitful applications to plasma physics \cite{EbFoFi17,EbFoFi20,Kelbg63,Kelbg6318}.

\begin{figure}[htbp]
\begin{center}
\includegraphics[height=7 cm,width = 6cm,angle=0]{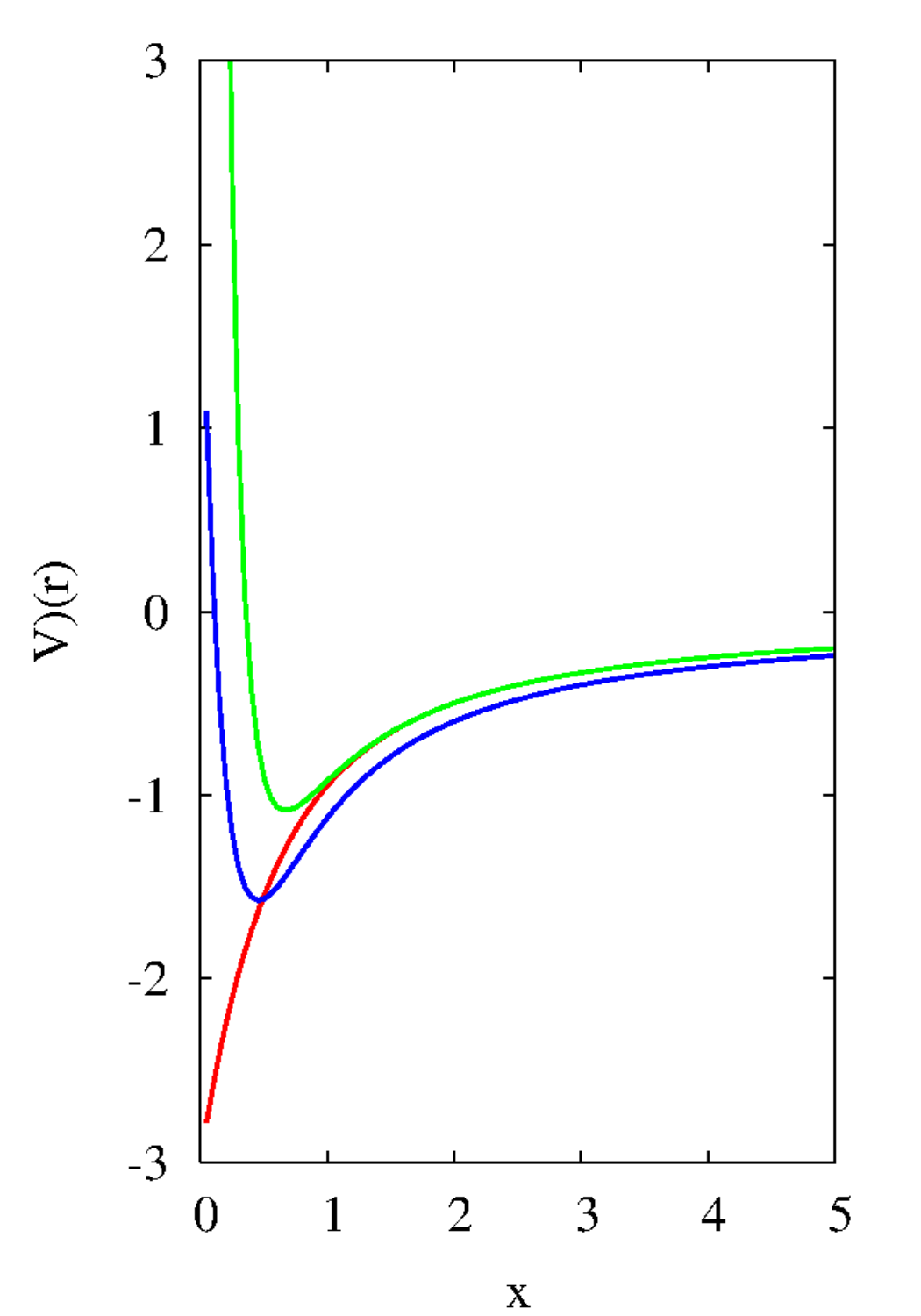}
\includegraphics[height=7cm,width=6cm,angle=0]{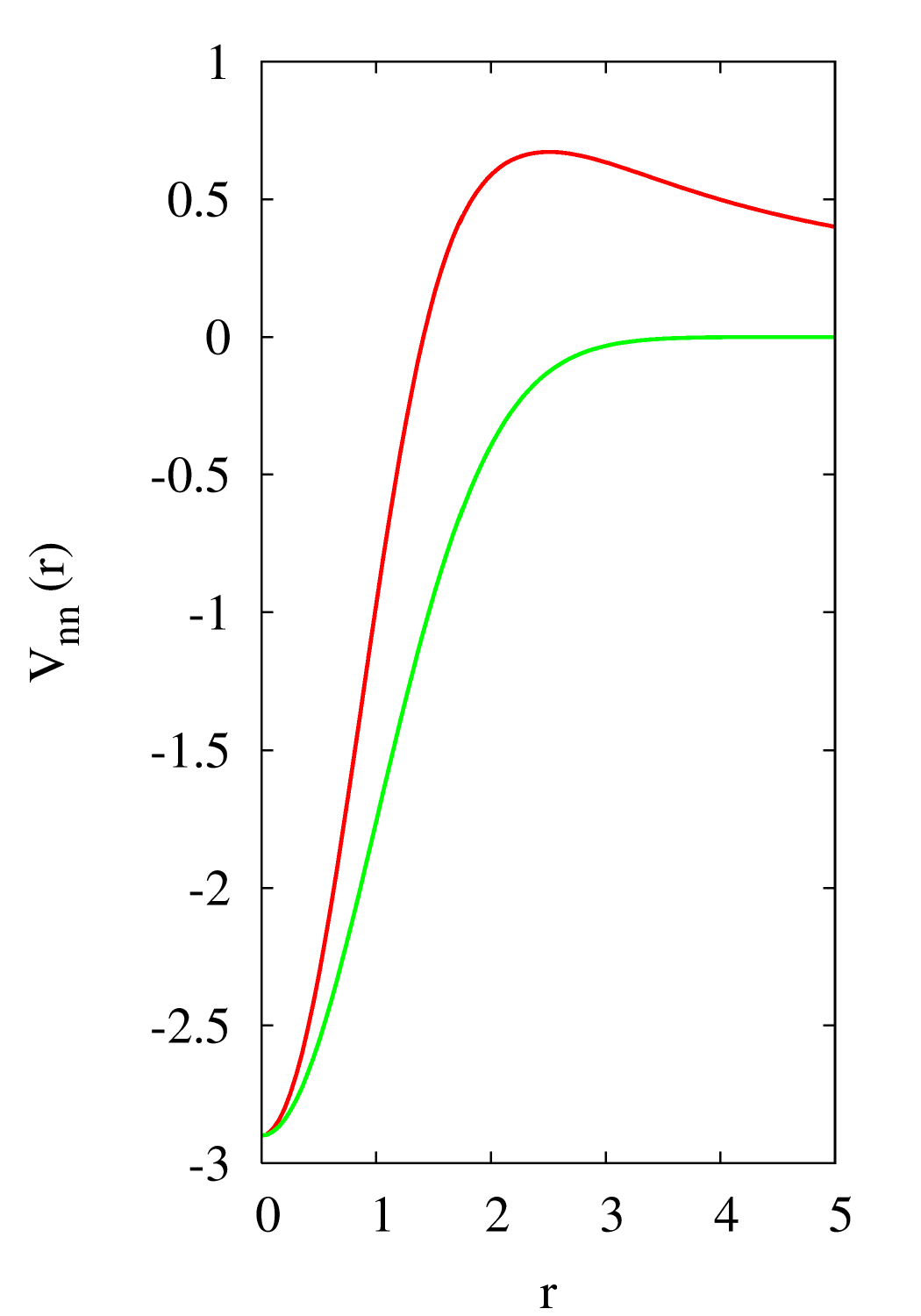}
\caption{(Colour online) Left-hand: alkali plasmas: exponential potential for opposite ions (red) for two different parameters of repulsive forces (in blue and green). 
Right-hand: high-temperature plasmas including reactions: exponential potential combined with attracting fusion forces (red curve). For comparison, the bare nuclear potential is shown (in green). The hump resulting from additional repulsive forces is the Coulomb barrier. Reactions may occur only by coming from the free states on the r.h.s. and crossing the barrier. The result of reactions are bound states close to $r=0$.  }
\label{fig-1}
\end{center}
\end{figure}

{\bf Additional short-range forces}: Now we introduce the second free parameter by adding a prefactor
to the exponential
\bea
V_{ij} (r) = \frac{e_i e_j}{4 \piup  \epsilon r} [1 - A_{ij} \exp(-\alpha r)].
\eea
This opens the door for new applications, in particular to alkali plasmas \cite{Hellmann1,Hellmann2,Hellmann3,Hensel,Iakubov,Richert}.  A good description of opposite charged particles in Cs-plasma provides, e.g., the choice $A=1.583$, \mbox{$\alpha = 1.082 \cdot 10^8$~cm$^{-1}$~{\cite{{EbFoFi17,EbFoFi20}}}}.

For many problems, the exponential potential, even in its extended form, is just a first approximation and additional short-range forces should be added, e.g., in the form
of additional short-range terms $V_{ij}' (r)$: 
\bea
\psi_{ij} (r) = \frac{e_i e_j}{4 \piup  \epsilon r} [1 - \exp(-\alpha r)] + V_{ij}'(r).
\eea
Several authors studied alkali plasmas  in this spirit \cite{Krasko,Richert}. Due to the 
complex structure of the ions we need here a more detailed description by more parameters \cite{Hensel,Iakubov}. Rather flexible is the addition of the following short-range exponential \cite{Krasko,EbFoFi17,EbFoFi20,Richert}
\bea
V_{ij}' (r) =  B_{ij} \exp \Bigg(- \frac{r^2}{R_{ij}^2} \Bigg), \qquad B_{ij} = \frac{e_i e_j}{4 \piup \epsilon} \frac{a}{R_{ij}}.
\label{eq-4}
\eea
The parameters $a$ and $R_{ij}$ can be considered as additional adjustable parameters. 
For the examples Cs and (Na) we find, e.g., \cite{EbFoFi17,EbFoFi20}
$a = 2.214, (3.362)$ a.u., $R=0.871, (0.487)$ a.u.. Note that the sorts indexes $i,j$ were here omited. Potentials of this type are nowadays generally accepted for the description of real 
plasmas, in particular for alkali plasmas \cite{Sadykova}. Applications to electrolytes have also been proposed~\cite{EbKrCMP23}. So far extended exponentials are not widely used in electrolyte theory
in comparison with hard-core models. However, one should not overlook that on the one hand charged hard cores 
are very successful in modelling the electrolytes, but on the other hand all quantum-chemical calculations lead 
to exponential-like potentials as was  already known by Hellmann \cite{Hellmann1,Hellmann2,Hellmann3}. On the other hand the simulations of the mean 
force potential  show that the effective interactions of ions in solution are rather complicated but always smooth \cite{Krienke13}.
We will also discuss here  a class of problems connected with fusion plasmas. The fusion of nuclei in the center of the sun and in modern fusion devices is due to attracting nuclear forces \cite{Atzeni,Lindl}. In reality, fusion plasmas contain many species of charged particles as electrons and several nuclear species like protons $p$, deuterons $d$, tritium nuclei $t$ and helium nuclei He$^{++}$. In the first approach to fusion plasmas we leave the electrons out by smearing them in a background and restrict ourselves to just 
one nuclear reaction between $p$ and $d$, neglecting also helium nuclei, which is essenetial for the so-called 
$p$-$p$-chain keeping only one pair reaction
having a relatively high gain:
\bea
p + d \rightarrow  t + 5.43~\text{MeV}.
\eea
Further, we do not consider here the spin as a variable and use another frequently used simplification  
by smearing out the electrons, so that they
form a uniform negative background. In order to model the internucleon energetic 
processes, we use the simple short-range potential:
\bea
V_{ij}'(r) =  - A \exp \big( - a_0 r^2\big).
\eea
The constants for the equations are phenomenological, that is, they are determined by fitting the equations to experimental data in a way that the properties of nucleon-nucleon interaction are modelled at least qualitatively, i.e., $A \sim 5$ MeV, $a_0 \sim 3~\text{fm}^{-2}$.

For the illustration in  figure~\ref{fig-1} we  schematically present the form of exponential potential for opposite ions including two different parameters of repulsive forces. The applications which we discuss here are just examples and we do not claim to develope
an exhaustive theory. Anyhow, we want to demonstrate here again  
that the exponential potential introduced into the statistical theory of charged particles
by Glauberman and Yukhnovskii in Lviv and and Kelbg in Rostock,  
is a quite rich potential model with respect to applications to electrolytes and plasmas. 

\section{Classical Coulombic systems}
\subsection{Charge distributions}
The basic characteristic quantity which we will study here, i.e., the charge-charge correlations in Coulombic systems,
which may be expressed by the pair correlation function which is defined in the Bogolyubov approach
as  \cite{FalkenhagenEb71,Falkenhagen,Holovko}
$$
F_{ij} (1,2) = 1 + g_{ij} (1,2)
$$ 
with the correlation function defined in the Bogolyubov theory by
\bea
g_{ij}(1,2) + \beta V_{ij}(1,2) + \beta \sum_k n_k \int V_{ik}(1,3) g_{jk} (2,3)\, \rd {\bf 3} = 0.
\eea
This approximation may be considered as the first term of an  expansion with respect to the plasma parameter \cite{FalkenhagenEb71,Falkenhagen,Holovko}.
We introduce now the charge density around a charge of species $i$ by
\bea
\sigma_i (r) = \sum_j n_j e_j g_{ij} (r).
\eea
The charge density has in average the opposite sign of the central ion, which  
determines many physical quantities as thermodynamic functions, structure factor etc.

As shown by Yukhnovskii and Kelbg, a simple and effective model of an electrolyte is the 
exponential interaction with only one $\alpha$-parameter. The charge density is for this model 
exponentially decaying at small densities $2 \kappa/\alpha <1$, ($\kappa=1/r_{\rm D}$, $r_{\rm D}$ is the Debye radius) \cite{Holovko},
\bea
\sigma_i (r ) = - \frac{e_i \kappa^2}{4 \piup \epsilon r (p^2 - s^2)} \big[\exp(-pr) - \exp(-sr)   \big]
\label{eq-9}
\eea
with \cite{FalkenhagenEb71,Falkenhagen,Holovko}
\bea
 \frac{p}{\alpha} = \frac{1}{2}\left( \sqrt{1 + 2 \frac{\kappa}{\alpha}} -\sqrt{1- 2 \frac{\kappa}{\alpha}}\right), 
 \qquad \frac{s}{\alpha} = \frac{1}{2} \left(\sqrt{1 + 2 \frac{\kappa}{\alpha}} +\sqrt{1 - 2 \frac{\kappa}{\alpha}}\right).
 \label{eq-10}
\eea
In the case that $\alpha$ is species-dependent, the solution is more complicated but for small $\kappa$ we may use $p = \kappa$ and $s = \alpha_{ij}$. 
The Coulomb energy is determined by the given charge density around an ion and we find
the result  \cite{Holovko} 
\bea  
U_C =  \sum_i N_i {U^{c}_{i}} = - \frac{V}{8 \piup }  \frac{\kappa^3}{\sqrt{1 + 2 \kappa/\alpha}} 
= - \frac{V}{8 \piup \epsilon} \frac{\kappa^3}{(1 + 2 \gamma/\alpha)}.
\eea
Here, $\gamma$ is a variable defined by $\kappa = 2 \gamma(1 + \gamma/\alpha)$ which provides a form of the equations, which is convenient for comparisons with the Debye--Huckel~(DH) and the mean spherical approximation (MSA theory) \cite{Holovko}.
For  large densities with $2 \kappa/\alpha >1$, i.e., $2 /\alpha  > r_{\rm D}$,  we find a solution 
which is oscillating~\cite{Holovko}:
\bea
&&\sigma_i (r) =   \textcolor{purple}{-} e_i \frac{\alpha^2}{8 \piup \epsilon qt}  \big[\exp(-qr) \sin(t r)
\big],\\ 
&&t/\alpha = (1/2) \sqrt{2 (\kappa/\alpha) -1}, \quad q/\alpha = (1/2) \sqrt{1 + 2 \kappa/\alpha}.
\eea
At  $\kappa > \alpha/2$, the charge density is in the Yukhnovskii--Kelbg (YK) theory given by
\bea
\sigma_i^{\rm YK} = - e_i  \frac{2 \kappa^2}{ \sqrt{4 (\kappa/\alpha)^2 -1}} \bigg[\exp(-qr) \frac{\sin(tr)}{r}\bigg].
\eea
The transition from exponential to oscillatory behavior occurs at $2 \kappa/\alpha = 1$. We note further that the result for the electrical energy for $2 \kappa/\alpha >1$ is the same as for lower  values of $\kappa/\alpha$, i.e., in this order there is no qualitative 
change of thermodynamics at the point of transition.

\subsection{Thermodynamic functions}
Standard models for the thermodynamic functions of classical ionic solutions are known since the work by Debye, H\"uckel and Bjerrum in 1920-ies based on the model of hard charged spheres. These models are well elaborated \cite{FalkenhagenEb71,Falkenhagen,Yukhn80,Yukhn25}. Problems of particular interst were modern versions of 
Bjerrum's mass action law and the question of Coulombic phase transitions \cite{Grigo,Grigo82,Fisher}.
 Surprisingly enough, some questions are still open for the  
class of exponential potentials in spite of so many studies \cite{Yukhn80,Yukhn25}.
Here, we plan to fill this evident gap by giving some further clasical and quantum statistical calculations.
The first result is about the virial series for exponential potentials including the second virial coefficient. 
Without giving the needed calculations which are rather trivial in detail, we give the virial series 
for exponential potentials.

A perturbation theory may be based on the quasi-exponential approximation \cite{Schmitz,Schmitz2}
\bea
F_{ij} (r) = \exp(g_{ij})  = (1  + g_{ij}) + \big[ \exp(- \beta V_{ij}' + g_{ij}) - 1 - g_{ij} \big],
\eea
which leads to a cluster expansion for the free energy \cite{FalkenhagenEb71,Falkenhagen,EbFoFi17,EbFoFi20,EbCMP25,Schmitz2}
\bea
F= F_{\rm id} + F_{\rm exp} +  \frac{V}{2} \sum_{ij} n_i n_j \int \rd {\bf 2} \bigg[\exp(-\beta V_{ij}' + g_{ij} ) - 1 - g_{ij} - \frac{1}{2} g_{ij}^2 \bigg] 
+ \cdots. 
\eea
The first correction to the ideal part is here given by the exponential interaction part which corresponds to an improved Debye law.
The success of the description by exponential potentials is based on the fact, that an essential part of the short-range effects is
included by an optimal choice of the exponential potential.
According to Yukhnovskii and Kelbg, we get for the contribution of the exponential potential~\cite{FalkenhagenEb71,Falkenhagen} 
\bea
&&\beta F_{\rm exp} = \frac{\kappa^3}{12 \piup} {\tilde f} \left(\frac{\kappa}{\alpha}\right), \qquad  
{\tilde f}\left(\frac{\kappa}{\alpha}\right) = \frac{3 \alpha^3}{4 \piup \kappa^3 }\left(Y^2 - 2 Y + \frac{2}{3} Y^3 - \frac{1}{2}Y^4 + \frac{5}{6}\right), \nonumber \\
&&Y = 2 + X, \qquad X = \sqrt{1 + 2 \frac{\kappa}{\alpha}}.
\label{ringfct}
\eea
The excess chemical potential is given by expression~\cite{FalkenhagenEb71,Falkenhagen} 
\bea
\mu_i^{\rm ex}  = - \frac{e_i^2 \alpha}{4 \piup \epsilon} \frac{\sqrt{1 + 2 \kappa/\alpha - 1}}{\sqrt{1 + 2 \kappa/\alpha}}=- \frac{e_i^2 \alpha}{4 \piup \epsilon} \bigg(1 - \frac{1}{\sqrt{1 + 2 \kappa/\alpha}} \bigg).
\eea
Because of the complicated analytical structure, a simple analytical Debye--H\"uckel-like approximation is often used 
\bea
\mu_i^{\rm ex1}  = - \frac{e_i^2 \kappa}{4 \piup \epsilon} \frac{1}{1 + 1.5 \kappa/\alpha}. 
\label{eq-19}
\eea
This first low-order contribution due to the exponential potential have been extensively discussed.
The relation between the Debye--H\"uckel-like (DH), which corresponds to ionic size $a={3}/{2 \alpha}$, and the YK-function for the chemical potential was discussed in detail in \cite{EbCMP25}. Here, it has been shown that essential differences appear in the asymptotic behaviour for large $\kappa^*$ ($\kappa^*=\kappa \sqrt{\alpha_{e}}$, where $\alpha_{e}$ is the degree of association).
For the pressure we may also use  an approximation which is correct only in the first deviations from Debyes law
and correct in the asymptotic for large $\kappa$, for example:
\bea
&&\beta p \simeq n_0^* + 2 n^* - \frac{\kappa^*3}{12 \piup} \varphi_{\rm YK} \bigg(\frac{\kappa^{*}}{\alpha} \bigg), \nonumber\\
&&\varphi_{\rm YK} (x) \simeq \frac{1}{ 1+ (3/8) \sqrt{\piup} x + (\piup/6) x^2} ,
\label{eq-20}
\eea
where $ n_0^*= n_0(1-\alpha_{e})$, $n^*=n \alpha_{e}$, $x=(\kappa*/\alpha)$.

\begin{figure}[t]
	\begin{center}
		\includegraphics[scale=0.31]{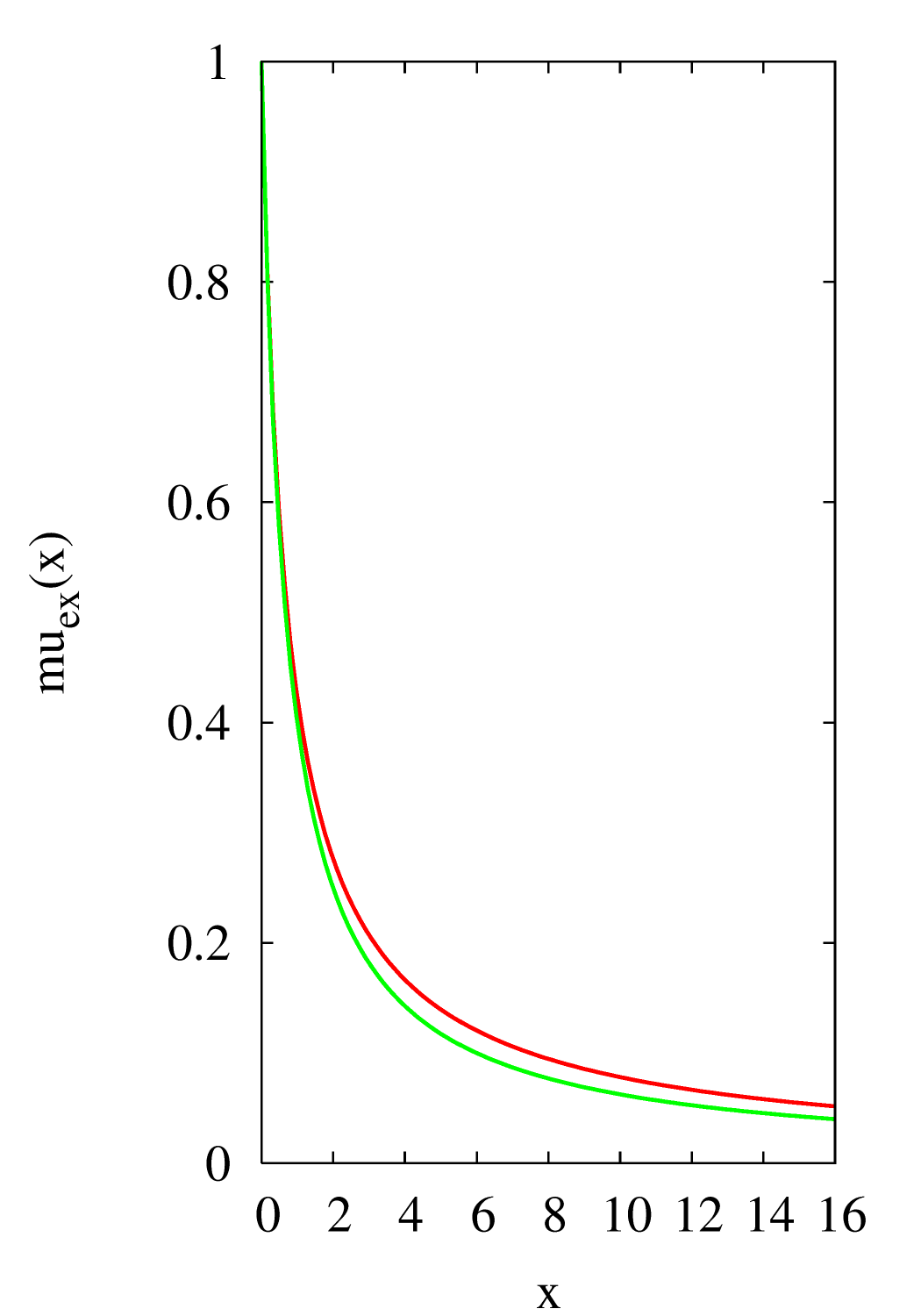}
		\caption{(Colour online) The excess chemical potential (red) and a Debye-H\"uckel-like aproximation (green) which is 
			exact only for small densities but nicely reproduces the overall shape. }
	\end{center}
	\label{fig-2}
\end{figure}

We calculate now the contribution of bound states to the second virial coefficient in the  form of a series, like it is known for hard core systems. We concentrate here on symmetrical binary systems which is an enormeous
simplification and most essenstial results remain also true  for unsymmetrical systems, at least asymptotically~\cite{FalkenhagenEb71,Falkenhagen}. For the contributions of forth and higher orders in the interactions, 
we define in the case $B=0$, the contribution of the 4th and higher orders by:
\bea
( \delta F)_{4+} =  \frac{V}{2} \sum_{ij} n_i n_j \int \rd {\bf 2} \bigg[\exp(-\beta V_{ij} ) - 1 + \beta V_{ij} - \frac{1}{2} (\beta V_{ij})^2 -  \frac{1}{6} (\beta V_{ij})^3  \bigg] ,   \label{eq-21} 
\eea
we remember that $B$ is the parameter of short-range interaction~\eqref{eq-4}.
For the symmetrical case, there are many compensations and it occurs that the most essential contributions are provided by the terms which are even in $(e_i e_j)$ and define a dimensionless characteristic function known from the hard-core case~\cite{FalkenhagenEb71,Falkenhagen}. For the exponential potential, the characteristic function reads including an asymmetric Gaussian repulsion:
\bea
&&M(b_e,B,R) = \frac{1}{2} \int_0^{\infty} \rd x\, x^2 \exp\left\{B \exp\left[-b \exp\left(-\frac{r^2}{R^2}\right)\right]\right\}\left\{\exp\left[ \frac{b_e}{x} \left(1 - \exp(-x)\right)\right]\right. \nonumber\\
&& \left.- \exp\left[-\frac{b_e}{x} \left(1 - \exp(-x)\right)  -2 - \frac{b_e^2}{x^2} \left(1 - \exp(-x)\right)^2\right]\right\}, \quad b_e =Z^2  \ell \alpha, \quad \ell = \frac{e^2}{4 \piup  \epsilon}, 
\eea
where $Z$ is the valency of ions in the considered symmetric system, $\ell \longrightarrow  e^2 / (4 \piup  \epsilon kT)$, $T$ is the temperature.
Here, $b_e$ is  for the exponential potential a modified Bjerrum parameter, and the characteristic bound state function $M(b_e)$ has for large $b_e$ the asymptotics:
\bea
M(b,B=0) \simeq {\sqrt{\frac{\piup}{8}}} \frac{\exp(b_e)}{b_e^{3/2}}.
\eea
Note that this asymptotics is different from that of the Bjerrum function~\cite{FalkenhagenEb71,Falkenhagen}. The Taylor expansion of the bound state function reads in terms of $b_e$: 
\bea
M (b_e,B=0) = \sum_{m=2,4,..} C_{2m} b_e^{2m}.
\eea
For the first coefficient $C_4$, we get, e.g.,
\bea
C_4 = \int_0^{\infty} \frac{\rd x}{24 x^2} \big[1 - \exp(-x)\big]^4 \simeq \frac{1}{48}.
\eea
We know that for hard charged spheres, the corresponding Bjerrum function $M(b)$ possesses a quickly convergent Taylor series with known coefficients \cite{FalkenhagenEb71,Falkenhagen}.
For exponential potentials, the other coefficients $C_{2m}$ are still to be calculated.
The result for hard spheres may be taken as a first rough approximation for the case of exponential interactions. 
In the case of extended exponential potentials, we have additional contributions to be bound state function.

\subsection{Bound states, mass action laws and phase transitions}

As shown in \cite{FalkenhagenEb71,Falkenhagen} the function $m(b)$ known as Ebeling function which is really a part of the second virial coefficient which, being not included in the linear theory, determines, in the case of hard charged spheres, 
the mass action constant of association and this way determines the bound state effects.
For the exponential potential, we get for associating electrolytes with $b_e \gg 1$,
the mass action function 
\bea
K(T) = \frac{8 \piup}{\alpha^3} M(b_e,B,R), \qquad b_e = \frac{Z^2 e^2 \alpha}{4 \piup \epsilon kT}  .
\eea
We see that $b_e$ plays for exponential interactions the role of the old Bjerrum interaction parameter $b = Z^2 e^2 a / 4 \piup \epsilon kT$. We note that our result refers so far to the
case that there are no additional interactions $B_{ij} =0$. 
In the case of additional attractive terms with large $B$-parameters, possibly a minimum at finite $r$ is formed as for alkali plasmas. In this case, more advanced,
possibly quantum-statistical calculations are needed. 
The similarity of the classical case to the 
theory of symmetrical associating ionic solutions analyzed in \cite{Falkenhagen,EbZpC71,Grigo,Grigo82} is striking.
The nonideal mass action law for the degree of association $\alpha_e$  of an exponential system reads 
\bea
\frac{1-\alpha_e}{\alpha_e^2} = \frac{8 \piup n_i}{\alpha^3} \exp(-\beta \mu_{\rm ex}^{\pm}) M(b_e,B,R),
\label{eq-27}
\eea
where $\mu_{\rm ex}^{\pm} = \mu_{\rm ex}^{+} + \mu_{\rm ex}^{-}$ is the total  excess chemical potential of the solution. 
As an application, we consider now the question of thermodynamic stability.
In lowest approximation we use the expression~\eqref{eq-19}, i.e., the Debye--H\"uckel-like lowest approximation 
for the excess chemical potential which allows analytical checks of the stability \cite{EbFoFi17,EbFoFi20}.
One of the necessary conditions for thermodynamic stability reads
\bea
\partial \mu_{\rm ex}^{\pm} / \partial n >0.
\eea
An ionic solution or a plasma is stable, if the total excess chemical potential, the sum of contributions of both charges  is increasing \cite{EbZpC71}, otherwise the charges will move against the concentration gradient
(diffusion instability).  At the border between stability and instability we have critical parameters. For the first approximation to the chemical potential given by equation~\eqref{eq-19}, the problem may be solved analytically  similar to  \cite{EbFoFi17,EbFoFi20}, and  we find in the present case that the border between stability and instability is reached at 
\bea
b_e^{cr} = \frac{32}{3}, \qquad (\kappa \ell)^{cr} \simeq 16.
\label{eq-29}
\eea
At stronger interactions and at higher densities, the system shows diffusion instabilities and forms two phases with different properties. For hard sphere ions, this was first shown  in \cite{EbZpC71}, and was analyzed in detail later, e.g., in \cite{Fisher,Stell}.
This analysis also includes  numerical and experimental confirmations~\cite{Schroer}. 
We also refer here  to more precise estimates for ionic hard core solutions 
and several experimental studies~\cite{Fisher,Schroer}. 
It would be interesting to check the influence of the deviations of the YK-theory of the chemical potential from the DH-theory at large $\kappa^*$ \cite{EbCMP25} and further to check the critical density for the new theory
of exponential potentials by simulations. Evidently, simulations for exponential potentials are not yet known.

\section{Statistical theory of quantum plasmas} 
\subsection{Generalized Kelbg potential  and pair ditribution}
\label{sec-3.1}
All the known classical results can be transferred to the quantum case using Kelbs recipe which provides a reasonable first approximation. Kelbg did not show in a fundamental work that for point charges, i.e., for hydrogen, the quantum interactions have approximately the form of the classical 
exponential interactions \cite{Kelbg63,Kelbg6318,EKK76}. This way the classical theory for the exponential potential leads 
in a first approximation directly to approximate results for quantum plasmas. We have just to replace 
the classical $\alpha$ which is a constant, through the thermal wave length, which is temperature-dependent
\bea
\frac{1} {\alpha} \rightarrow \lambda = \frac{{\bar h}}{\sqrt{\mu k_{\rm B} T}} .
\eea
Here, $\mu$ is the relative mass of two particles which meet. Using Kelbgs recipe, we may transfer all the known classical results to the
quantum case, which provides at least a reasonable first appoximation. The Kelbg approximation provides the correct physics in a very simple way. The strict quantum statistics, however,
is much more difficult. As demonstrated first by Kelbg for the pure Coulomb potential, i.e., for hydrogen, the strict quantum theory may be developed based on Slater functions or like Vedenov-Larkin and others on Green functions \cite{EbFoFi20,EKK76}. 
So far the Kelbg theory has not been generalized to other models of interactions as, e.g., the exponential potential.
We will very briefly give here only the lowest approximations for the exponential potential 
and discuss higher approximations in the interaction parameter and the role of bound states 
only,  assuming a weak degeneracy. We will show that this extension is of some relevance for alkali plasmas
and possibly  for fusion plasmas as well.
The main application of a quantum statistics of exponential interactions is so far to alkali plasmas which are often  descibed by the exponential interaction model \cite{Krasko,EbFoFi17,EbFoFi20,Sadykova}.

Systems with exponential interactions may be treated in a similar way as Yukawa systems, which are a prototype of interactions having a Fourier transform.
The Coulomb case appears as a limiting case and there are many other interesting applications. 
As well known, the transition from classical to quantum systems may be most easily performed 
by using the technique of Slater functicons, which for a pair of charges $a$, $b$ are denoted as $S_{ab}(r)$
and are a kind of generalization of the pair Boltzmann factor
\bea
\exp(-\beta V_{ab} (r)) \rightarrow S_{ab}(r).
\eea
We use representations of the Slater function by plane waves as already used in the pioneering papers by Uhlenbeck and Bethe  \cite{EKK76,EbFoFi17,EbFoFi20} and find:
\bea
 S_{ab} (r) = \mbox{A} \int \rd {\bf k} \exp[- {\bf k}\cdot {\bf r}] \exp[-\beta \hat{H}_{ab}] \exp[+\ri{\bf k} \cdot {\bf r}], \quad
{\bar h} H_{ab} = -\frac{\bar h}{2m_{ab}}\Delta + V_{ab}, \nonumber
\eea
where the first term is the kinetic part of hamiltonian, $m_{ab}$ is the relative mass.
The Slater function of  a pair is the diagonal element of the pair density operator. We may use  the general
expansion technique for density operators with $e^2$ as a small interaction parameter. This way we get
the expansion 
$$
S_{ab} = S_{ab}^{(0)} + S_{ab}^{(1)} + S_{ab}^{(2)} + \dots \,.
$$
 The Fourier representation in the distance coordinate space reads
\be
{\tilde S}^{(n)}_{ab}({\bf t})  = \int \rd {\bf r}\,
S^{(n)}_{ab}({\bf r}) \exp[ \ri {\bf t} \cdot {\bf r}].
\ee
In order to find the virial coefficient we need only an integral over the Slater sum and it is sufficient to get the Fourier transforms in the zero point $t = 0$.
In order to calculate the Fourier transforms of the Slater function we define
\bea
v_{ab}({\bf r}) = \exp[- \ri {\bf k} \cdot {\bf r}] \exp[-\beta \hat{H}_{ab}]
\exp[\ri {\bf k} \cdot {\bf r} ], \nonumber
\eea
and the Fourier transform
\begin{equation}
 v_{ab}({\bf k},{\bf t}) = \int \rd{\bf r} \exp(\ri {\bf t} \cdot {\bf r}) v_{ab} ({\bf k},{\bf r})  .
\end{equation}
Using these definitions we are able to use our general perturbation scheme by transforming to the $v_{ab}$-functions.
We find for the first two iterations
\bea
v_{ab}^{(1)}({\bf k},{\bf t}) &=& - \beta \int_0^{\beta} \rd \beta' \,{\tilde V}_{ab}({\bf t}) \exp[-\lambda_{ab}^2 ({\bf t}^2 - 2 {\bf k} \cdot {\bf t})
(1 - \beta' / \beta)],  \nonumber\\
v_{ab}^{(2)}({\bf k},{\bf t})& = & \frac{1}{8 \piup^3} \int_0^{\beta} \rd \beta' \int_0^{\beta'} \rd \beta'' \int \rd {\bf t'}\, {\tilde V}_{ab} ({\bf  t} - {\bf t'}) {\tilde V}_{ab} ( {\bf t'})\nonumber\\ 
&&\exp\{ -\lambda_{ab}^2 [({\bf t}^2 - 2 {\bf k}\cdot {\bf t} )(1 - \beta'/\beta) + ({\bf t'}^2 - 2 {\bf k}\cdot {\bf t'} )(\beta'/\beta - \beta''/\beta)]\} .
\eea
On the basis of these approximations for the Bloch functions, we find the
free energy up to the second order
in the coupling strength $e^2$. In the same way, we may go further to any order.
By introducing the first iteration of the $v_{ab}$ function, which is the Kelbg approximation,  given before into the formula for the Slater functions, for the corresponding Fourier transforms negelecting symmetry effects we get 
\bea
{\tilde S}_{ab} ({\bf t}) = \frac{\lambda_{ab^3}}{\piup^{3/2}} \int \rd {\bf k} \exp(- \lambda^2 {\bf k}^2)
 v_{ab}({\bf k},{\bf t},\beta).
\eea
Similar expressions for Coulomb and Yukawa plasmas are already known~{\cite{EbFoFi17,EbFoFi20}} and may be  combined. This way we get the Slater function in first approximation, which contains three Fourier-transformable contributions
\bea
{\tilde S}_{ab}^{(1)}({\bf t}) = - \beta 4 \piup \ell_{ab} \left(\frac{1}{t^2} - \frac{1}{t^2 + \alpha^2}\right)   
M\left(1,\frac{3}{2}; -\frac{1}{4}\lambda_{ab}^2 t^2\right)  + B_{ij} F_{ab}^G (\bf{ t}),
\eea
where the first terms are expressed by the Kummer function (or confluent hypergeometric function) $M(a,b;z)$
similar to the Coulomb potential in Kelbgs theory. 
The last term corresponds to the Gaussian contribution, which is still to be extra evaluated
but may be possibly better included into the short range contributions,
\bea
F_{ab}^G (\mathbf {t}) = - \beta \piup R_{ab}^3  \int_0^{\beta} d \beta' \exp\bigg[- \frac{{\bar h}^2}{2 m_ab} 
({\mathbf t}^2) - 2 ({\mathbf k \cdot \mathbf t}) (\beta - \beta')  - \frac{\beta'}{4} R_{ij}^2 {\mathbf t}^2  \bigg] .
\eea
Introducing the Fourier transform of the Gaussian, we get
\bea
{\tilde S}_{ab}^{(1)}({\bf t}) = - \beta \left[ 4 \piup \ell_{ab}  \left( \frac{1}{t^2} - \frac{1}{t^2 + \alpha^2}\right)
+   \piup B_{ab}  R_{ab}^3  \exp\left( - \frac{\beta}{4} R_{ab}^2 {\bf t}^2 \right) \right] M\left(1,\frac{3}{2}; 
-\frac{1}{4}\lambda_{ab}^2 t^2\right) .
\eea
The evaluation of the inverese Fourier transform is still to be done.
We have now a closed expression for the Fourier transform of the unscreened distribution function including exponential and Gaussian interactions.  In some cases it may be useful to shift the last term $F_{ab}^G$ 
to the short-range terms. The inverse transformations are not yet known.
By using the aproximation $M(a,b;z) \sim 1/[1 - (a/b) z]$ we get for the long-range part the simple expression 
\bea
&&{\tilde S}_{ab}^{(1)}({\bf t}) = - \beta 4 \piup \ell_{ab} \left( \frac{1}{t^2} - \frac{1}{t^2 + \alpha^2}\right)
\cdot \frac{1}{(1 + \lambda_{ab}^2 t^2 /6)}   \simeq - \beta 4 \piup \ell_{ab} \left(\frac{1}{t^2} - \frac{1}{t^2 + \alpha_q^2} \right) , \\
&&\frac{1}{\alpha_q^2} = \frac{1}{\alpha^2}  + \frac{\lambda_{\pm}^2}{6}. 
\label{eq-40}
\eea
Note that this quite simple result for alkali plasmas reduces for quantum hydrogen 
with $\alpha \rightarrow \infty$ to Kelbgs known estimate $\alpha_q =\sqrt{6}/\lambda$.
Since the formulae have now the same form as for the classical case, we may repeat
the classical YK-screening procedure in the same way by just replacing $\alpha \rightarrow \alpha_q$,
and get this way new quantum pair correlation functions for alkali plasmas.
Performing then the Fourier inverse transform leads for $\kappa < \alpha_q /2$ to 
\bea
g_{ij} ({\bf r}) = - \frac{ \ell_{ij}}{(p_q^2 - s_q^2) r}  \big[ \exp(- p_q r) - \exp(-s_q  r) \big] 
\label{eq-41} 
\eea
and for larger plasma densities with $\kappa > \alpha_q /2$ it leads to an oscillatory correlation function
\bea
&&g_{ij} ({\bf r}) = -  \frac{2 \ell_{ij} \kappa^2}{ \sqrt{4 (\kappa/\alpha_q)^2 -1}} \bigg[\exp(-q_q r) \frac{\sin(t_q r)}{r}\bigg], \label{eq-42}\\
&&t_q = (\alpha_q/2) \sqrt{2 (\kappa/\alpha_q) -1}, \quad q_q = (\alpha_q/2) \sqrt{1 + 2 \kappa/\alpha_q}.
\eea
We see that in the present approximation the form of the classical YK-formulae is retained with replacing 
the classical $\alpha$ by the new characteristic quantum parameter $\alpha_q$ defined by equation~\eqref{eq-40}.
This corresponds to the approximation of the Kummer function given before. 
We note here that formulae for the quantum parameters of alkali plasmas similar to equation~\eqref{eq-40} 
had been used already before as in an empirical
extension of the theory of alkali plasmas including quantum effects \cite{EbFoFi17,EbFoFi20}. Here, the quantum parmeters for alkali plasmas have been derived in a consequent way from the quantum statistics of 
the exponential potential. A more precise evaluation of the integrals which is still to be done would include error functions like in the original Kelbg theory for pure Coulomb systems.

\subsection{Free energy, pressure and mass action law}
Having access to the Fourier transform of the correlation functions [here ${\tilde S}_{ab}({\bf t})$] opens the way to several applications to quantum systems.  For instance, Bogolyubovs quantum statistics for Bose gases demonstrate that contributions to the excess pressure, and the sound velocity $c_s$ can be represented by Fourier transforms of the two particle correlations
\bea
\delta (\beta p) = n^2 {\tilde S}_2 ({\bf t} = 0), \quad c_s = \sqrt{n  {\tilde S}_2 ({\bf t} = 0) }.
\eea
Similar to Bogolyubov's theory, for plasmas contributions to the excess free energy due to the non-Coulombic parts of the interactions in terms of the Fourier transforms of the Slater functions we get
\bea
\delta F =  - \frac{1}{2} V \sum_{ab} n_a n_b \cdot
\sum_{p \geqslant 1} \Big[\delta {\tilde S}_{ab}^{(p)} ({\bf t} = 0)\Big] .
\label{FreeEnergy}
\eea
Using these formulae we get corrections to the Coulombic part of the free energy and may also obtain the sound velocity. In a similar form we may derive    corrections to the pressure of the plasma
\bea
\delta (\beta p) = \sum_{ab} n_a n_b \cdot
\sum_{p \geqslant 1} \left[{\tilde S}_{ab}^{(p)} ({\bf t} = 0)\right] .
\label{Pressure}
\eea
For additional potentials in the form of exponential or Gaussian potentials, we do not have divergence problems in carrying out the integrals. Note that $t \rightarrow 0$ corresponds to $r \rightarrow \infty$. 
Putting together these results we arrive at expressions for the free energy of 
the gases with exponential Coulombic interactions. So far in this consideration we neglected the exchange effects for the first two orders in the coupling $g_{ab}$. 
Note, that the ``quantum virial functions'' $Q$ for the direct contributions are in principle infinite series in the coupling 
strength $e^2$. We calculated here only the first lower orders. 

We are coming now to the problem of quantum bound states. 
We study here the theory for exponential interactions only in 
an elementary quasi-chemical approach. Further we restict the consideration to regions
of still weak correlations and low density. This is of relevance in particular for the studies
of alkali plasmas~\cite{EbFoFi17,EbFoFi20,Richert,Sadykova} and possibly also for fusion plasmas,
as will be shown. In a first approximation, we may follow the observation of Kelbg,
that the classical theory for the exponential potential leads directly to approximate results for hydrogenic quantum plasmas
by replacing the classical $\alpha$ which is a constant expressed by the quantum wave length, which is temperature-dependent. In the case of the exponential potential, the Kelbg prescription is to be replaced 
by 
\bea
\frac{1} {\alpha} \rightarrow \frac{1} {\alpha_q}  = \sqrt{ \frac{1}{\alpha^2}  + \frac{\lambda_{\pm}^2}{6}}.
\eea
The new quantum $\alpha$-parameter $\alpha_q$  depends on the temperature and on $\mu_{\pm}$ as the relative mass of two particles (electron and ion or proton and deuteron), which meet and 
possibly may form bound states. However, we have to note here that so far our screening procedure is valid only for the case when we approximate all relative masses by just one, namely that for the most relevant pair as a rule $\mu_{\pm}$ or the reacting pair, i.e., $\mu_{pd}$. 
In this way, the potential between the reacting particles is possibly strongly influenced by screening.

Using Kelbgs recipe we may transfer all the known classical results to the
quantum case which provides a reasonable first appoximation. This refers, in particlular, to the excess chemical potential which determines the mass action law, the degree of ionization $\alpha_e$. This way instead of equation~\eqref{eq-27} we get the quantum Saha equation for exponentially interacting Coulomb systems
\bea
\frac{1-\alpha_e}{\alpha_e^2} = n K_2 (T) \cdot \exp[- \beta \ell G_{\rm YK} (\kappa^*,\lambda, \alpha_q)].
\label{eq-47}
\eea
Here is $\alpha_e = n^*/n$, the relation of density of free charged particles to the total density of that species
(free and bound); the screening quantity $\kappa^*$ is to be calculated as $\kappa^* = \kappa \sqrt{\alpha_e}$. 
An aproximate quantum version of the ionization depression function $G_{\rm YK}(x)$ was developed recently in \cite{EbCMP25}.
The new approximation corresponding to the present more complete approximation equation~\eqref{eq-42} has still to be worked out. 

 The new function for the exponential potential called $Q_{\rm YK}$-approximation (QYKA) 
is similar to the standard Debye--H\"uckel like approximation (QDHA) but it is characterized by a longer range
with respect to the density-dependent screening parameter $\kappa^*$.
In the plasma theory based on the Yukhnovskii--Kelbg approach, we use a temperature-dependent  
quantum $\alpha$-parameter $\alpha_q$, and the dimensionless parameters $x = \kappa \lambda_{\pm}$, 
$X = \sqrt{1 + 2 \kappa/ \alpha_q}$ as well as the Yukhnovskii--Kelbg-functions:
\bea
&&G_{\rm YK} (x) = \frac{6}{\sqrt{\piup} x} \left[ 1 - \frac{1}{\sqrt{1 + (\sqrt{\piup}/3) x}} \right], \qquad x = \frac{\kappa}{\alpha_q}, \label{eq-49}\\
&&\varphi_{\rm YK}(x) = \frac{1}{x^3} \left( \frac{1}{4} X^3 - \frac{2}{3}X +2 - \frac{3}{4X} \right) = 1 -  \frac{9}{4} x + \frac{9}{2} x^2 - \frac{35}{4} x^3 \dots \,.
\label{eq-50}
\eea 
This way
we get the equation of state (EOS) for the plasma 
\bea
\beta p  = n_i^* + n_e^*  + n_0^*   - \frac{\kappa^{*3}}{24 \piup} \varphi_{\rm YK} \big(\kappa^*,\lambda,\alpha_q \big), 
\qquad \kappa^{*2} = 4 \piup \ell Z^2 (n_e^* + n_i^*),
\eea
with the mass action law 
\bea
\frac{n_0^*}{n_e^* n_i^*} =  8 \piup \sqrt{\piup} \lambda_{ie}^3 \sigma (T) 
\exp \big[- Z^2 \ell \kappa^* G_{\rm YK}(\kappa^* , \lambda, \alpha_q) \big].
\eea
This is all we know so far about the pressure and the IPD (ionization potential depression) of the plasmas with exponential interaction. We discuss now the mass action constant of the assumed pair reactions and assume in this simplest version of the theory that a bound state spectrum $E_{sl}$ is known or can be estimated by solution of the Schr\"odinger equation. This way we arrive at the expression
\bea
K_2 (T) =   \sum_{sl} g_{sl} [\exp(- \beta E_{sl}) - 1 + \beta E_{sl}]. 
\eea 
Following Planck and Larkin, we assume here the standard renormalization procedure, which is used to avoid singularities \cite{EbRoPOP26}. From earlier work we know that in a strict quantum-statistical theory, the convergence is achieved in this smooth, analytical way \cite{EbRoPOP26}. For the Coulomb potential, the quantum-statistical  exact treatment of the bound state problem is known \cite{EbFoFi17,EbFoFi20,EbRoPOP26}. For the exponential potential, this problem is still to be solved
in a strict way similar to \cite{EbRoPOP26} for helium. 
Staying here within an elementary approach, we need only estimates for the bound state spectrum and the mass action law. We  should notice that this simple version of the theory neglects all finer effects as, e.g., the interaction between charged and neutral particles \cite{Hensel}.

\subsection{Applications to alkali plasmas and to fusion plasmas}
The model of plasmas interacting by exponential potential is quite general which we  demonstrate here by giving two examples. The first one is devoted to alkali plasmas \cite{EbFoFi17}.

\begin{figure}[h]
\begin{center}
\includegraphics[scale=0.31]{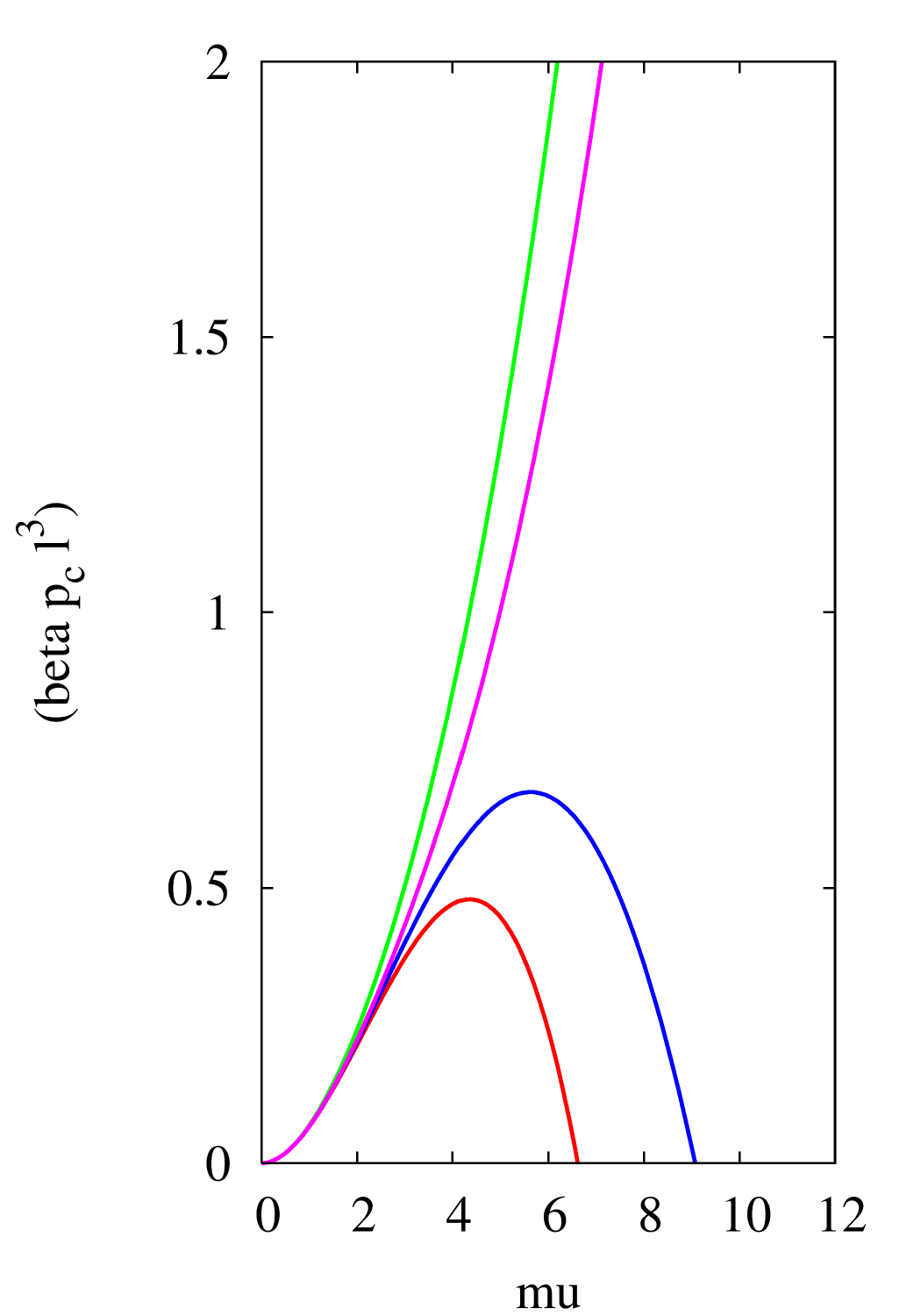}
\caption{(Colour online) Qualitative shape of the high-temperature isotherms of the pressure function $\pi_c =p_c \ell^3/k_{\rm B} T$ of alkali plasmas for different temperatures as a function of the Bogoliubov sreening parameter $\mu_B = \ell \kappa^*$ [we used equation~\eqref{eq-55} for the drawing with $1/\alpha = 2$~\AA~for $T = 1000, 3000, 6000, 10000$~K (from below)]. According to our estimate, the curves for $T < 3000$ K are unstable and the critical situation 
is a bit above $T \sim 3000$ K which is at least in the region of observed critical points for alkali plasmas. }
\label{fig-3}
\end{center}
\end{figure}

{\bf Alkali plasmas:} We first define a dimensionless pressure function by defining $\pi(\mu_B) =p \ell^3/ k_{\rm B} T$.
This quantity depends mainly on the Bogolyubov plasma parameter $\mu_B = \ell \kappa^*$ and the temperature, but scrupulously also on the characteristic material parameter $\alpha$ and on the temperaure through $\lambda$. In figure~\ref{fig-3}  we show the pressure function of an alkali-like plasma which is described by a simplified EOS like equation~\eqref{eq-21} over $\mu_B$ for different parameters in a schematic way. The critical Bogoliubov parameter $\mu_c \simeq 8$ is for our model alkali plasma only about one half of the value for hydrogen \cite{EbFoFi17}.

In order to simplify the calculations of the pressure, we used a simplified Pade formula~\eqref{eq-21}
for the dimensionless Coulomb part of the pressure using the approximation  and 
find this way an expression which is not more complicated than the Van der Waals equation
\bea
\pi_c = \frac{p_c \ell^3}{k_{\rm B} T} = 0.0796 \mu_B^2 - 0.02653 \frac{\mu_B^3}{[1 + A  (\mu_B/a) + B (\mu_B/b)^2]} .
\eea 
We should note however that this Pade approximation is only qualitatively correct, since it takes into account 
only the first deviations from the limiting law and the aymptotics by adapting the parameters~$A,B$.
The drawing in figure~\ref{fig-3} is a principal scheme for alkali-like plasmas made with equation~\eqref{eq-55} with adapted parameters $A, B$ is an estimate. Accordingly, the shape of the pressure is only semi-quantitative,
although it shows a reasonable behaviour including a phase-transition-like behaviour with wiggles below an estimated critical temperature about $T \sim 3000$ K. We remember that the experiments for cesium, rubidium, natrium, potassium and lithium plasmas show a gas-liquid transition in the region $T_c \simeq 2000$--3500~K~\cite{Hensel}. However, our curves are not quantitative, 
or theoretical chritical temperature is still too high in comparison with experimental data, which is $T_c \simeq 2000$ K for cesium and  for other alkali plasmas is in the range up to 3500 K \cite{Hensel}.
One of the reasons of the still existing disagreement between the present theory and the 
experiments is that we neglected higher order contributions in the formula~\eqref{eq-49} and~\eqref{eq-50},  as well as polarization and other relevant effects \cite{EbFoFi17,Hensel}. 
This way, the theory of alkali plasmas given here still needs  several refinements, an advantage being, however, the complete analyticity of the formulae in the framework of the exponential potential \cite{Falkenhagen,Yukhn25}. 

{\bf Proton-deuteron fusion plasmas}: The following approach to a  model of fusion plasmas is to be considered just as a first approach which has to be checked and worked out.  In order to avoid any misunderstandings we would like to say first, that we do not have in mind here controlled thermonuclear fusion plasma, which are generally highly non-equilibrium
but have in mind natural fusion plasmas like in the Sun which are probably in a kind of local
thermodynamic equilibrium. Here, the term local thermodynamic equilibrium means that some degrees of freedom are 
in equilibrium and a local temperature and a local pressure exist. The degrees of ionization are not in equilibrium  
and the study of this degree of ionization is our interest. This is a traditional object of astrophysics \cite{Salpeter,Sturrock}.
However, one may hope that any progress in this field might be also beneficial  for the study of inertial fusion in laboratories.
In order to model fusion plasmas in local thermodynamic equilibrium, we use here the model discussed 
in section~\ref{intro}  in a simplified version based on the Kelbg approximation as in subsection~\ref{sec-3.1}. We think it is needless to say here once again that due to increasing needs of energy sources, the physics of the largest natural ressource of energy is to be further explored. The fusion processes in the center of the Sun and maybe sometimes regarding the technical fusion equipment are nowadays in the center of interest. We mention only a few of recent fusion studies \cite{Atzeni,Lindl}.
We want to prove here in a small study that the exponential potential may be useful  in modelling a fusion plasma in 
local thermodynamic equilibrium. So far we used the exponential potential already for modeling different kinds of modifications of the Coulomb law at small distances. In fusion plasmas we have a new kind of force, the nuclear force which is responsable for energy providing reactions like the so-called $p$-$p$-chain reaction which plays a big role in the Sun and has a relatively high gain. 

\begin{figure}[t] 
	\begin{center}
		\includegraphics[scale=0.3]{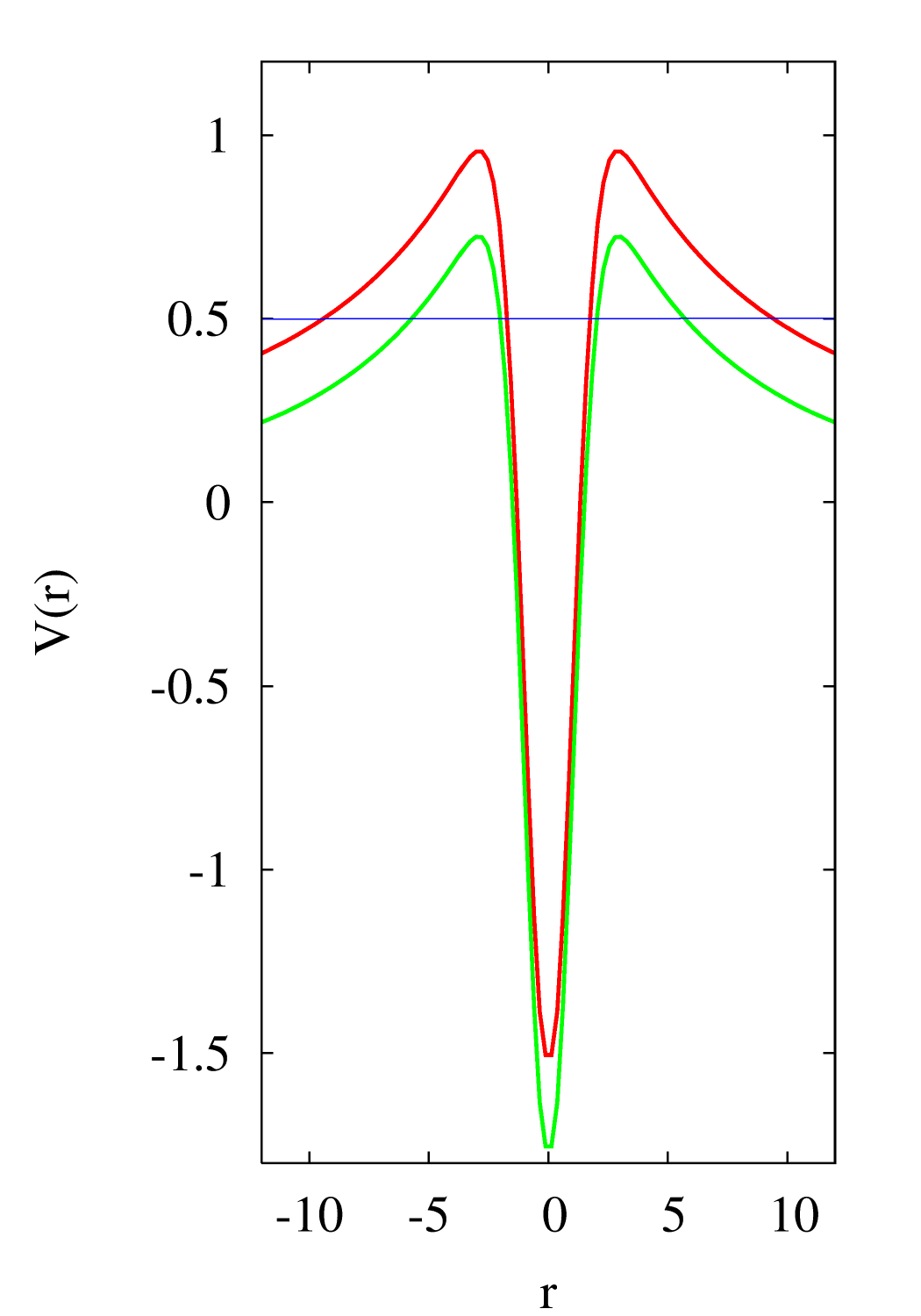}
	\end{center}
	\caption{(Colour online) Exponential potential model equations~\eqref{eq-4} and~\eqref{eq-56} for repulsing Coulomb and attracting nuclear forces between 
		two charged nuclei (in red). The bound nuclear states are located within the well, only the highest ones may be 
		involved in tunneling processes. The screened potential  is also shown (in green) for higher densities
		as well as the hypothetically assumed energetic level of the free plasma particles (blue line at 0.5). 
		Evidently, the tunneling through the barrier is supported by screening~\cite{Salpeter}, which was treated here in the YK-approximation which is much better 
		than previously applied screening procedures.
	}
	\label{fig-4}
\end{figure}

The nature of nuclear forces is complex. The discussion of these problems is beyond the scope of this study. We will confine ourselves to use a very simple model of the nucleon-nucleon interaction
which fulfils just the condition that the ground state is at least in first approximation in agreement with the observations \cite{Salpeter}. 
Our simple ansatz consists of an exponential Coulomb term and a contribution of fusion forces. This ansatz allows us to model the hump between the Coulomb tail and the well of attracting nuclear forces:
\bea
V_{nn} (r) = \frac{e^2}{4 \piup \epsilon r} [1 - \exp(\alpha_0 r)]  - A \exp \big( - a_0 r^2\big).
\label{eq-55}
\eea
Here, $1/\sqrt{a_0}$ characterizes the range of the nuclear attractive forces ($a_0$ is a characteristic parameter of nuclear forces)  
and $1/\alpha_0$ is the range where the validity of the Coulomb law ends, both lengths beeing in the femtometer range (see the schema figure~\ref{fig-1}). 
The values of these two lengths can only be roughly estimated. We prefer values which are in accdordance with
the observed reaction gains.
The long range contribution to the interaction forces is modified in the plasma by screening effects.
By using the quantum-statistical version of the YK-screening theory given here we find for the screened potential
\bea
{\tilde V}_{nn} (r) = \frac{e^2}{4 \piup \epsilon r} \big[\exp(-p_q r) - \exp(s_q r)  - A \exp( - a_0 r^2)\big],
\label{eq-56}
\eea
where the parameters $p_q$, $s_q$ are given as before in the framework of the Yukhnovskii--Kelbg theory by 
equation~\eqref{eq-10} with the modification $\alpha \rightarrow \alpha_q$. A comparison of unscreened and screened potential between proton and deuteron is shown in figure~\ref{fig-4}.

The proper statistical definition of reaction constants in systems with Coulomb interacrtions 
is a quite difficult problem and the answer is not unique. 
According to Onsager, the definition of a bound species is not unique. We have some freedom in the choice of the definition {\cite{EbFoFi17,EbFoFi20}. In the case of a Coulombic pair, an alternative good choice is according to Planck, Larkin and others \cite{EbRoPOP26}
given by the prescription to omit the divergent terms in a Taylor expansion. 
We show in figure~\ref{fig-4}, how the assumed profile of the actual Coulomb and nuclear potential 
is changed by screening the exponential part of the potential in a dense plasma according to the given formulae for the screened potential \eqref{eq-9} and \eqref{eq-10}  modified by quantum effects according to equation~\eqref{eq-47} for $\alpha_q r_{\rm D} > 2$. We also note  the transition to an oscillatory pair distribution
for $\alpha_q r_{\rm D} < 2$.

The quantitative description of the nuclear force relies on equations that are partly empirical \cite{Salpeter,Sturrock,Atzeni,Lindl}. 
Despite the long history of nuclear physics, it is still not possible to write down, exact formulae for the two-nucleon interaction. In a first approach we leave the electrons out by smearing them in a background and restrict ourseleves to just one nuclear reaction, also neglecting  helium nuclei in the $p$-$p$-chain keeping only one pair reaction having a relatively high gain.

\section{Conclusions}
This work is a continuation of the previous paper \cite{EbHoCMP26} where we aimed to extend the work by Yukhnovslii and Kelbg on the classical and quanten statistical theory of the exponential potential. We show that this theory is 
very powerful with respect to new applications. Here we develop, e.g., new applications to electrolytes and alkali plasmas and also propose  a new application to fusion plasmas. By using and extending the old analytical results for the thermodynamic functions of Coulomb systems with exponential interactions, we succeed to give
applications to such complicated problems like phase transitions in electrolytes and alkali plasmas 
by using fully analytical and simple graphical methods. We underline this way the power of analytical theories
which should not be neglected in comparision with new powerful simulation methods,
since they offer a more comprehensive understanding of the details.  

\section*{Acknowledgement}
We express our thanks to several colleagues for suggestions, providing additional material and encouragement,
in particular to Ronald Redmer and Gerd R\"opke from the Rostock University for discussions about the
quantum-statistical theory and the applications. We thank Oksana Patsahan from Yukhnovskii Institute for Condensed Matter Physics for the discussions of phase transition in ionic systems. We also express our sincere thanks to referees which contributed much to clarify our aims and to improve our work.

\newpage
\ukrainianpart

\title{До статистичної теорії сильних електролітів і високотемпературної
	плазми: нові застосування праць Юхновського та Кельбга ІІ}
\author{В. Ебелінг\refaddr{label1}, М. Головко\refaddr{label2}}
\addresses{
	\addr{label1}{Інститут фізики університету Гумбольдта, Берлін, Німеччина}
	\addr{label2}{Інститут фізики конденсованих систем ім.~І.~Р.~Юхновського Національної академії наук України, вул.~Свєнціцького,~1, 79011 Львів, Україна}
	}

\makeukrtitle

\begin{abstract}
	\tolerance=3000%
	Експоненціальні потенціали використовувалися з часів робіт Крамерса, Гельмана, Глаубермана, Юхновського та Кельбга для розв’язання задач квантової хімії, теорії іонних розчинів і плазми. В даній роботі ми далі розвиваємо ці підходи та представляємо нові результати, зокрема стосовно квантово-статистично теорії.
	Зокрема, отримано квантовий потенціал Кельбга для експоненціальних взаємодій і обговорюється екранування, а також проблема термодинамічної стійкості. Далі наводяться нові застосування якісної теорії плазми на основі лужних елементів та плазми на основі термоядерного синтезу. Показано, що використання експоненціального потенціалу та попередніх аналітичних результатів з праць Юхновського та Кельбга уможливлюють якісний аналітичний розгляд таких складних проблем, як фазові переходи в електролітах та плазмі на основі лужних елементів.
	\keywords сильні електроліти, високотемпературна плазма, осциляційні кореляції
	
\end{abstract}

\lastpage

\begin{thebibliography}{10}

\bibitem{Hellmann1}
Hellmann~H., J. Chem. Phys., 1935, \textbf{3}, 61, \doi{10.1063/1.1749559}.

\bibitem{Hellmann2}
Hellmann~H., Acta Physicochim. URSS, 1935, \textbf{1}, 913, (in German).

\bibitem{Hellmann3}
Hellmann~H., Acta Physicochim. URSS, 1936, \textbf{4}, 324, (in German).

\bibitem{GlaYuk52_1}
Glauberman~A.~E., Yukhnovskii~I.~R., Zh. Eksp. Teor. Fiz., 1952,
\textbf{22}, 562--572, (in Russian).

\bibitem{Yukhn54}
Yukhnovskii~I.~R., Zh. Eksp. Teor. Fiz., 1954, \textbf{27}, 690--698, (in Russian).

\bibitem{Kelbg59}
Kelbg~G., Wiss. Z. U. Rostock MNR, 1959/60, \textbf{9}, 41.

\bibitem{Kelbg62}
Kelbg G., In: Electrolytes, Pesce B. (Ed.), Pergamon Press, New York, London, 1962, 109.

\bibitem{FalkenhagenEb71}
Falkenhagen~H., Ebeling~W., In: Ionic Interactions: From Dilute Solutions to
Fused Salts, Vol.~1, Petrucci~S.~(Ed.), Academic Press, New York, 1971,
1--59, \doi{10.1016/B978-0-12-553001-9.50006-2}.

\bibitem{Falkenhagen}
Falkenhagen H., Theorie der Elektrolyte, Hirzel, Leipzig, 1971, (in German).

\bibitem{Krasko}
Krasko~G.~L., Gurskii~Z.~A., JETP Lett., 1969, \textbf{9}, 363.

\bibitem{EbFoFi17}
Ebeling W., Fortov V. E., Filinov V., Quantum Statistics of Dense Gases and Nonideal Plasmas, Springer Seriesin Plasma Science and Technology, Springer International Publishing, Cham, 2017, \doi{10.1007/978-3-319-66637-2}.

\bibitem{EbFoFi20}
Fortov V. E., Filinov V. S., Larkin A. S., Ebeling W., Statistical Physics of Dense Gases and Nonideal Plasmas, FizMatLit, Moscow, 2020, (in Russian).

\bibitem{Yukhn80}
Iukhnovskii I. R., Golovko M. F., Statistical Theory of Classical Equilibrium Systems, Naukova Dumka, Kyiv, 1980, (in Russian).

\bibitem{Yukhn25}
Yukhnovskii I. R., Holovko M. F., Statistical Theory of Classical Equilibrium Systems, 2nd ed., Akademperiodyka, Kyiv, 2025, \doi{10.15407/akademperiodyka.558.444}.

\bibitem{EbHoCMP26}
Ebeling~W., Holovko~M., Condens. Matter Phys., 2026, \textbf{29}, 23501, \doi{10.5488/cmp.29.23501}.

\bibitem{Hansen}
Hansen~J.-P., McDonald~I.~R., Theory of simple liquids: with applications to
soft matter, Academic Press, Elsevier, Oxford, Amstersdam, fourth edn., 2013,
\doi{10.1016/C2010-0-66723-X}.

\bibitem{Richert}
Richert~W., Insepov~S.~A., Ebeling~W., Ann. Phys. (Berlin, Ger.), 1984, \textbf{496}, No.~2, 139--150, \doi{10.1002/andp.19844960207}.
\bibitem{EbZpC71}
Ebeling~W., Z. Phys. Chem., 1971, \textbf{247}, 340,  \doi{10.1515/zpch-1971-24741}.

\bibitem{Grigo}
Ebeling~W., Grigo~M., Ann. Phys. (Leipzig, Ger.), 1980, \textbf{37}, 21, \doi{10.1002/andp.19804920104}.

\bibitem{Grigo82}
Ebeling~W., Grigo~M., J. Solution Chem., 1982, \textbf{11}, 151,
\doi{10.1007/BF00667599}.

\bibitem{Fisher}
Fisher~M.~E., Levin~V., Phys. Rev. Lett., 1993, \textbf{71}, 3826,
\doi{10.1103/PhysRevLett.71.3826}.

\bibitem{Stell}
Stell~G., J. Stat. Phys., 1995, \textbf{78}, 197,
\doi{10.1007/BF02183346}.

\bibitem{Schroer}
Schr\"oer~W., J. Mol. Liq., 2011, \textbf{164}, 3,
\doi{10.1016/j.molliq.2011.08.003}.

\bibitem{Kelbg63}
Kelbg~G., Ann. Phys. (Berlin, Ger.), 1963, \textbf{467}, 219--224, \doi{10.1002/andp.19634670308}, (in German).

\bibitem{Kelbg6318}
Kelbg~G., Ann. Phys. (Berlin, Ger.), 1963, \textbf{467}, 354--360, \doi{10.1002/andp.19634670703}, (in German).

\bibitem{EKK76}
Ebeling~W., Kraeft~W.~D., Kremp~D., Theory of Bound States and Ionization
Equilibrium in Plasmas and Solids, Akademie-Verlag, Berlin, 1976.

\bibitem{EbCMP25}
Ebeling~W., Condens. Matter Phys., 2025, \textbf{28}, 23101,
\doi{10.5488/CMP.28.23101}.
\bibitem{Hensel}
Redmer~R., Hensel~F., Holst~B. (Eds.), Metal-to-Nonmetal Transitions,
Springer, Berlin, 2010, \doi{10.1007/978-3-642-03953-9}.

\bibitem{Holovko}
Holovko~M., In: Proceeding of Shevchenko Scientific Society, Vol.~XXIX,
Collected Physical Papers, 2011, \textbf{8}, 452--467, (in Ukrainian).

\bibitem{Schmitz}
Schmitz~G., Phys. Lett., 1966, \textbf{21}, 174, \doi{10.1016/0031-9163(66)90304-0}.

\bibitem{Schmitz2}
Ebeling~W., Kelbg~G., Schmitz~G., Ann. Phys. (Leipzig, Ger.), 1966, \textbf{473}, No.~1--2, 29--41, \doi{10.1002/andp.19664730105}.

\bibitem{Iakubov}
Alekseev~V.~A., Iakubov~I.~T., Phys. Rep., 1983, \textbf{96}, 1,
\doi{10.1016/0370-1573(83)90074-1}.

\bibitem{Sadykova}
Sadykova~S.~P., Ebeling~W., Tkachenko~I.~M., Eur. Phys. J. D, 2011,
\textbf{61}, 117, \doi{10.1140/epjd/e2010-10118-y}.

\bibitem{EbKrCMP23}
Ebeling~W., Krienke~H., Condens. Matter Phys., 2023, \textbf{26}, 23602,
\doi{10.5488/CMP.26.23602}.

\bibitem{Krienke13}
Krienke~H., Condens. Matter Phys., 2013, \textbf{16}, 43006, \doi{10.5488/CMP.16.43006}.

\bibitem{EbRoPOP26}
Ebeling~W., R\"opke~G., Plasma Phys., 2026, \textbf{33}, 032705, \doi{10.1063/5.0313273}.

\bibitem{Salpeter}
Salpeter~E.~E., Aust. J. Phys., 1954, \textbf{7}, 373,
\doi{10.1071/PH540373}.

\bibitem{Sturrock}
Sturrock~P.~A. (Ed.), Physics of the Sun: Volume I --- The Solar Interior, Springer Netherlands, 1986.

\bibitem{Atzeni}
Atzeni~S., Meyer-ter-Vehn J., The Physics of Inertial Fusion: BeamPlasma Interaction, Hydrodynamics, Hot Dense Matter, Oxford University Press, 2004, \doi{10.1093/acprof:oso/9780198562641.001.0001}.

\bibitem{Lindl}
Lindl~J.~D., Amendt~P., Berger~R., Glendinning~S., Glenzer~S.~H.,
Haan~S.~W., Kauffman~R.~L., Landen~O.~L., Suter~L.~J., Phys. Plasmas,
2004, \textbf{11}, 339--491, \doi{10.1063/1.1578638}.

\end{thebibliography}
\end{document}